\documentclass[aps,jcp,preprint,superscriptaddress]{revtex4}

\usepackage[utf8]{inputenc}

\usepackage{amsmath, amssymb}
\usepackage{tabularx,booktabs,array,dcolumn}
\usepackage{setspace}
\usepackage{vmargin}
\usepackage{graphicx,psfrag,subfigure}
\usepackage{color}
\usepackage[usenames,dvipsnames]{xcolor}

\usepackage[all]{xy}

\setmarginsrb{1.5cm}{2cm}{1.5cm}{1cm}{0cm}{0cm}{0cm}{1.5cm}
\newcommand{\mx}[1]{\mathbf{#1}}

\newcommand{\dd}{\mathrm{d}}

\newcommand{\cm}{cm$^{-1}$}

\def\dd{\mathrm{d}}
\def\Eh{E$_\mathrm{h}$}
\def\com{\mathrm{CM}}
\def\tr{^\mathrm{T}}
\def\massratio{m_a/m_b}

\def\cp{{P}}

\def\iGamma{ {\it\Gamma} }

\newcommand{\npart}{n_\text{p}}

\newcommand{\mr}[1]{\mathrm{#1}}

\def\mapart{m_a} 
\def\mbpart{m_b} 
\def\massratio{\mapart/\mbpart}

\def\X1Sgp{X\ ^1\Sigma_\mr{g}^+}
\def\B1Sup{B\ ^1\Sigma_\mr{u}^+}
\def\b3Sup{b\ ^3\Sigma_\mr{u}^+}
\def\a3Sgp{a\ ^3\Sigma_\mr{g}^+}
\def\e3Sup{e\ ^3\Sigma_\mr{u}^+}
\def\c3Pup{c\ ^3\Pi_\mr{u}^+}
\def\Sp{S_\mr{p}}
\def\Se{S_\mr{e}}

\def\EF1Sgp{EF\ ^1\Sigma_\mr{g}^+}

\makeatletter
\def\@cite#1#2{$^{\mbox{\scriptsize #1\if@tempswa , #2\fi}}$}
\makeatother

\begin{document}

\title{%
  Pre-Born--Oppenheimer Molecular Structure Theory
}
\author{%
  Edit M\'atyus\\[0.cm]
  Institute of Chemistry, E\"otv\"os Loránd University, 
  Pázmány Péter sétány 1/A, Budapest, Hungary\\
}
\date{\today}

\begin{abstract}
\noindent%
In pre-Born--Oppenheimer (pre-BO) theory a molecule is considered as a quantum system as a whole, 
including the electrons and the atomic nuclei on the same footing.
This approach is fundamentally different from the traditional quantum chemistry 
treatment, which relies on the separation of the motion of the electrons and the atomic nuclei.
A fully quantum mechanical treatment of molecules has a great promise for 
future developments and applications.
Its most accurate versions may contribute to the definition of 
new schemes for metrology and testing fundamental physical theories; 
its approximate versions can provide an efficient theoretical description 
for molecule-positron interactions and, in general, 
it would circumvent the tedious computation and fitting of 
potential energy surfaces and non-adiabatic coupling vectors. 
To achieve these goals, the review points out important technical and fundamental 
open questions.
Most interestingly, the reconciliation of pre-BO theory with the classical chemistry knowledge 
touches upon fundamental problems related to the measurement problem of quantum mechanics. 
\end{abstract}

\maketitle

%
%
\clearpage
\section{%
  Quantum chemistry vs. quantum mechanics and chemistry? 
  \label{ch:qcqm}\label{ch:intro}}
\noindent%
We start with a historical overview of the chemical theory
of molecular structure and the origins of quantum chemistry, which is followed by 
a review of methodological details and applications of pre-Born--Oppenheimer theory.

\subsection{Historical background: Chemical structure, physical structure from organic chemistry experiments}
    \begin{center}
      \begin{minipage}{0.9\linewidth}
        \emph{%
        [T]he dominating story in chemistry of the 1860s, 1870s, and 1880s
        was neither the periodic law, nor the search for new elements, nor the early
        stages of the study of atoms and molecules as physical entities. It was
        the maturation, and demonstration of extraordinary scientific and technological
        power, of the ``theory of chemical structure'' ... 
        }
        \begin{flushright}
          \textsc{Alan J. Rocke} \\[-0.15cm]
          Image and Reality: Kekulé, Kopp, and the Scientific Imagination \\
          (The University of Chicago Press, Chicago and London, 2010) 
        \end{flushright}     
      \end{minipage}
    \end{center}
\vspace{1cm}
During the second half of the 19th century, the pioneering organic chemists 
generation---represented by Williamson, Kekulé, Butlerov, Crum Brown, Frankland, 
Wurtz, etc.---had explored an increasing number of chemical transformation 
in their laboratory experiments and 
worked towards the establishment of a logical framework for their observations.
The ``first chemist' conference'', held in Karlsruhe on 3 September 1860, 
resulted in an internationally recognized definition of the atomic masses. 
This agreement ensured that the same molecular 
formula was then used for 
the same substance in all laboratories around the world, and 
thereby opened the route to the successful development of the theory 
of chemical structure.
The development of the chemical theory has been surrounded by 
heated debates about what is reality and what is mere speculation. 
We note that contemporary physics (gravitation, electromagnetism) 
was not able to provide any satisfactory description for molecules. 
To give a taste of this exciting period, we reproduce a few extracts 
from Alan J. Rocke's chemical history book \cite{Rocke10}:
    \begin{itemize}
      \item
        Friedrich August Kekulé [von Stradonitz] (1858): 
        ``rational formulas are reaction formulas, and 
        can be nothing more in the current state of science.''
      \item       
        Friedrich August Kekulé [von Stradonitz] (1859):  
        ``[he] rejected the possibility of determining the physical arrangement of 
        he constituent atoms in a molecule from the study of chemical reactions, 
        since chemical reactions necessarily alter the arrangements of the atoms in the molecule.''
      \item
        Charles Adolphe Wurtz (1860): %
        ``[W]e do not have any means of assuring ourselves in 
        an absolute manner of the arrangement, or even the real existence of 
        the groups which appear in our rational formulas...  merely express parental ties.''
      \item
        Hermann Kolbe (1866): %
        ``Frankly, I consider all these graphical representations... 
        as dangerous, because the imagination is thereby given too free rein.''
      %
      \item
        Johannes Wislicenus (1869): %
        ``[it] must somehow be explained by the different arrangements of their atoms in space.''
      \item
        Jacobus Henricus van’t Hoff (5 September 1874, 1875): %
        ``La chimie dans l’espace''
      \item
        Joseph Achille Le Bel (5 November 1874): %
        physical structure in the 3-dimensional space
    \end{itemize}
    \vspace{0.15cm}

\subsection{Historical background: Application of quantum theory to molecules}
At the time when Erwin Schrödinger wrote down his famous wave equation \cite{Sch26},
the concept of the classical skeleton of the atomic nuclei arranged in 
the three-dimensional space 
was already a central idea in molecular science derived from the organic chemists' 
laboratory experiments.
The idea of a separate description of the electrons and the atomic nuclei, \emph{i.e.,}
the motion of the atomic nuclei on a potential energy surface (PES), 
which results form the solution of the electronic problem in 
the field of fixed external nuclear charges, is usually connected to 
the work of Born and Oppenheimer in 1927 \cite{BoOp27} and 
perhaps the later references \cite{Born51,BoHu54} are also cited. 
At the same time, in Ref.~\cite{SuWo13} Sutcliffe and Woolley analyze 
René Marcelin's doctoral dissertation published in 1914 (the author died during World War I), 
which appears to be the earliest work in which ideas reminiscent of 
a potential energy surface can be found.
They also argue that the idea of clamping 
the atomic nuclei in order to define an electronic problem was 
attempted already within the framework of The Old Quantum Theory, 
and later these attempts were taken over (more successfully) to 
the application of Schrödinger's theory for molecules. 

In any case, what we usually mean by quantum chemistry gains its equations 
from a combination of quantum mechanics and the Born--Oppenheimer (BO) approximation 
(and perhaps corrections to the BO approximation are also included). 
Sutcliffe and Woolley conclude that this theoretical framework results from
\emph{our} (chemically motivated) choice \cite{SuWo12} and: 
\emph{The tremendous success of the usual practice might perhaps be best regarded as 
a tribute to the insight and ingenuity of the practitioners for inventing 
an effective variant of quantum theory for chemistry.}

\subsection{Quantum chemistry}
The well-known theory applicable to molecules has grown out from 
the separation of the motion of the electrons and the atomic nuclei. 
This separation defines two major fields for quantum chemistry, 
electronic structure theory (EST) and the corresponding electronic Hamiltonian 
(in atomic units):
\begin{align}
  \hat{H}_\text{el}
  =
  -\sum_{i=1}^{n_\mr{e}}\frac{1}{2}\Delta_{\mx{r}_i} 
  +\sum_{i=1}^{n_\mr{e}}\sum_{j>i}^{n_\mr{e}} \frac{1}{|\mx{r}_{i}-\mx{r}_j|}
  -
  \sum_{i=1}^{n_\mr{e}}\sum_{n=1}^{n_\mr{n}} \frac{Z_n}{|\mx{r}_{i}-\mx{R}_n|}
  +
  \sum_{m>n}^{n_\mr{n}}\sum_{n=1}^{n_\mr{n}} \frac{Z_nZ_m}{|\mx{R}_{n}-\mx{R}_m|}
  \label{eq:elham}
\end{align}
with the $\mx{r}_i$ electronic and $\mx{R}_n$ nuclear positions and electric charges, $Z_n$;
and nuclear motion theory (NMT) with the Hamiltonian for the motion of the atomic nuclei
(or rovibrational Hamiltonian):
\begin{align}
  \hat{H}_\text{nuc}
  =
  \hat{T}(\rho) + \hat{V},
  \label{eq:nucham}
\end{align}
where $\hat{T}(\rho)$ is the rovibrational kinetic energy operator 
and the potential energy, $\hat{V}$, which is called the potential energy surface
and it is obtained from the eigenvalues of Eq.~(\ref{eq:elham})
computed at different positions of the atomic nuclei.

Several important chemical concepts gain a theoretical background from this separation, most
importantly the potential energy surface (PES) is defined.
Its minimum structure (or structures if it has several local minima) defines 
the equilibrium structure, which is a purely mathematical construct resulting 
from the separability approximation but it is usually identified with the 
classical molecular structure. 
Then, the nuclei are re-quantized to solve the Schrödinger equation
of the atomic nuclei, Eq.~(\ref{eq:nucham}), 
to calculate rovibrational states, resonances, reaction rates, etc.

The electronic structure and quantum nuclear motion theories 
have many similar features but each field have its own peculiarities. 
Most importantly, the spatial symmetries are 
different: in EST the point-group symmetry is defined by the fixed, 
classical nuclear skeleton, whereas in the PES depends only on the relative positions of the 
nuclei in agreement with 
the translational and rotational invariance of an isolated molecule.
Furthermore in EST, 
the molecular translations and rotations are separated off by fixing the atomic nuclei, 
and thus the kinetic energy can be written in a very simple form in Cartesian coordinates.
In NMT, it is convenient to define a frame fixed to the (non-rigid) body to separate off 
the translation and to account for the spatial orientation of 
this frame by three angles \cite{CSTSutcliffe,MHSutcliffe}. 
Thereby, in a usual NMT treatment, the coordinates are necessarily curvilinear. 
In spite of all complications, it was possible to develop automated 
procedures \cite{Lu00,LaNa02,YuThJe07,MaCzCs09,FaMaCs11}, 
which allow us to efficiently compute hundreds or thousands of rovibrational energy states 
for small molecules using curvilinear coordinates appropriately chosen for 
a molecular system
\cite{YuTeBaHoTi14,Te16,Ca17}.

\subsection{Quantum mechanics and chemistry?}
The direct treatment of molecules as few-particle quantum systems is much less explored. 
Nevertheless, we may think about molecules as a quantum system as a whole without 
any \emph{a priori} separation of the particles, 
which we call pre-Born--Oppenheimer (pre-BO) molecular structure theory. 
(This approach is also called non-Born--Oppenheimer theory 
in the literature \cite{CaBuAd03,chemrev13}.)

The $(\npart+1)$-particle time-independent Schrödinger equation
\begin{align}
  \hat{H}\Psi = E\Psi
  \label{eq:tise}
\end{align}
contains the non-relativistic Hamiltonian
\begin{align}
  \hat{H}
  =
  \hat{T}
  +
  \hat{V}, 
  \label{eq:ham}
\end{align}
which is
the sum of the kinetic energy operator
\begin{align}
  \hat{T}
  =
  -
  \sum_{i=1}^{\npart+1}
    \frac{1}{2m_i} \Delta_{\mx{r}_i}
  \label{eq:kin}    
\end{align}
and the Coulomb potential energy operator
\begin{align}
  \hat{V}
  =
  \sum_{i=1}^{\npart+1}\sum_{j>i}^{\npart+1}
    \frac{q_iq_j}{|\mx{r}_i-\mx{r}_j|}, 
  \label{eq:pot}
\end{align}
where atomic units are used and $\mx{r}_i$ labels 
the laboratory-fixed (LF) Cartesian coordinates of the $i$th particle associated with 
the $m_i$ mass and the $q_i$ electric charge ($i=1,2,\ldots,\npart+1$). 
The full molecular Hamiltonian 
has $2(\npart+1)$ parameters, the mass and the electric charge for each particle
$m_i$ and $q_i\ (i=1,2,\ldots,\npart+1)$, respectively. 
In addition, the physical solutions must satisfy the spin-statistics theorem, 
thereby the spins $s_i\ (i=1,2,\ldots,\npart+1)$ (the fermionic or bosonic character) 
appear as additional parameters. In total, there are $3(\npart+1)$ parameters, 
which define the molecular system. 
In addition, we may specify the quantum numbers corresponding to 
the conserved quantities of an isolated molecule: the total angular momentum, 
its projection to a space-fixed axis, the parity, and the spin quantum numbers 
labelled with $N$, $N_z$, $p$, $S_a$, $S_{z,a}$, $S_b$, $S_{z,b}$, $\ldots$ 
(for particle types $a$, $b$, etc.), respectively 
\footnote{We use the spectroscopists' notation \cite{GreenBook07} for 
the total angular momentum quantum number, $N$, instead of $L$ that is commonly used 
in the physics literature.}.

In a full, pre-BO treatment, the spatial symmetries of the few-particle quantum system 
are the same as in NMT. The potential energy is the simple Coulomb interaction term as in EST 
but now all electric charges belong to the quantum system. 
The choice of the coordinates for the kinetic energy operator 
is a question of convenience. We could use some appropriate curvilinear system 
to describe the rotating non-rigid body, similarly to NMT. 
The motivation for using laboratory-fixed Cartesian coordinates, 
similarly as in EST, is to develop a generally applicable theoretical and 
computational framework, similarly to the 
recent electron-nuclear orbital theories 
\cite{Nakai02,BoVaSh04,ChPaHS08,IsTaNa09,SiPaSwHS13,Lara13,lowdin13,CaChSuLi15}, 
which have grown out from EST with the aim of incorporating also (some of the) 
atomic nuclei in the quantum treatment and which can be generalized to 
the relativistic regime without having to deal with the cumbersome transformation 
of the operators to curvilinear coordinates.

Various direct and highly specialized solution techniques have been proposed 
in the literature \cite{BiCh81,Ko00,KaHi06,PaZiYe16} for the many-particle Schrödinger
equation, Eq.~(\ref{eq:ham}). 
Our approach, a variational solution method using 
explicitly correlated Gaussian basis functions (ECGs), 
is detailed in Section~\ref{ch:varsol}.
ECGs \cite{Ry03,rmp13} have been successfully used in EST \cite{KoBiVa12} and 
their application for molecules and in general for few-particle quantum systems 
has been pioneered by Adamowicz and co-workers \cite{CaBuAd03,chemrev13} and 
Suzuki and Varga \cite{SuVaBook98}.

%
%
\section{%
Variational solution of the electron-nuclear Schrödinger equation with 
explicitly correlated Gaussian functions
\label{ch:varsol}}
\noindent%
A general $(\npart+1)$-particle variational approach was developed in Refs.~\cite{MaRe12,Ma13} 
for the solution of the time-independent many-particle Schrödinger equation, Eq.~(\ref{eq:tise}),  
beyond spectroscopic accuracy \cite{spectracc} and 
for various combinations of the non-relativistic quantum numbers, 
$N,N_z,p,S_a,S_{z,a},S_b,S_{z,b},\ldots$. 
Our aim was to avoid any \emph{a priori} separation of 
the different particles, and thus the computational method is aimed to be applicable 
over the entire physically allowed range of the $3(\npart+1)$ physical parameters:
the $m_i$ mass, 
$q_i$ electric charge, and 
$s_i$ spin (bosonic or fermionic character) 
($i=1,\ldots,\npart+1$).

Details of the variational procedure are reviewed in the next subsections according to 
the following aspects:
\begin{itemize}
  \item[(1)]
    Coordinates: \\
    translationally invariant (TI) or laboratory-fixed (LF) Cartesian coordinates
  \item[(2)]
    Hamiltonian: \\
    TI and LF forms of the Hamiltonian
  \item[(3)]
    Basis functions: \\
    ECGs with a polynomial prefactor and adapted to the spatial symmetries
  \item[(4)]
    Matrix elements: \\
    analytic expressions with quasi-normalization and 
    pre-computed quantities using infinite-precision arithmetics
  \item[(5)]
    Eigensolver: \\
    direct diagonalization using LAPACK library routines, 
    non-orthogonal basis sets, numerical treatment of near-linear dependencies
  \item[(5+1)]
    Parameterization of the basis functions: \\
    the enlargement and refinement of the basis set with one function at a time,  
    fast eigenvalue estimator, sampling-importance resampling,  
    random walk or Powell's method for the refinement of the basis functions 
\end{itemize}

\subsection{Coordinates}
Translationally invariant (TI) and center-of-mass (CM) coordinates are obtained from
the LF Cartesian coordinates, $\mx{r}$, by a linear transformation
\begin{align}
  \left(%
    \begin{array}{@{}c@{}}
      \mx{x} \\
      \mx{R}_{\com} \\
    \end{array}
  \right)
  =
  (\mx{U}\otimes\mx{I}_3)\mx{r}
  \quad\Leftrightarrow\quad
  \mx{r}
  =
  (\mx{U}^{-1}\otimes\mx{I}_3)
  \left(%
    \begin{array}{@{}c@{}}
      \mx{x} \\
      \mx{R}_{\com} \\
    \end{array}
  \right),  
  \label{eq:ticoord}
\end{align}
where 
$\mx{x}$ is invariant upon the overall translation of the system, if 
the constant matrix $\mx{U}\in\mathbb{R}^{(\npart+1)\times(\npart+1)}$
has the following properties:
\begin{align}
  \sum_{j=1}^{\npart+1}U_{ij}=0, \quad i=1,\ldots,\npart 
  \quad\text{and}\quad
  U_{\npart+1,j}=m_j/m_\mr{tot}, \quad j=1,\ldots,\npart+1 .
  \label{eq:umx}
\end{align}
($\mx{I}_n$ denotes the $n\in\mathbb{N}$ dimensional unit matrix throughout this work.)
There are infinitely many possible TI coordinate sets, any two, 
$\mx{x}$ and $\mx{y}$,  of them are 
related by a linear transformation:
\begin{align}
  \left(%
    \begin{array}{@{}c@{}}
      \mx{y} \\
      \mx{R}_{\com} \\
    \end{array}
  \right)
  =
  (\mx{V}\mx{U}^{-1}\otimes\mx{I}_3)
  \left(%
    \begin{array}{@{}c@{}}
      \mx{x} \\
      \mx{R}_{\com} \\
    \end{array}
  \right)
  \quad\Leftrightarrow\quad
  \left(%
    \begin{array}{@{}c@{}}
      \mx{x} \\
      \mx{R}_{\com} \\
    \end{array}
  \right)
  =
  (\mx{U}\mx{V}^{-1}\otimes\mx{I}_3)
  \left(%
    \begin{array}{@{}c@{}}
      \mx{y} \\
      \mx{R}_{\com} \\
    \end{array}
  \right)
  \label{eq:coordtrfo}
\end{align}
with
\begin{align}
  \left(%
    \begin{array}{@{}c@{}}
      \mx{y} \\
      \mx{R}_{\com} \\
    \end{array}
  \right)
  =
  (\mx{V}\otimes\mx{I}_3)\mx{r}
  \quad\Leftrightarrow\quad
  \mx{r}
  =
  (\mx{V}^{-1}\otimes\mx{I}_3)
  \left(%
    \begin{array}{@{}c@{}}
      \mx{y} \\
      \mx{R}_{\com} \\
    \end{array}
  \right) .
  \label{eq:coordtrfo2}
\end{align}

\begin{figure}
  \includegraphics[scale=1.]{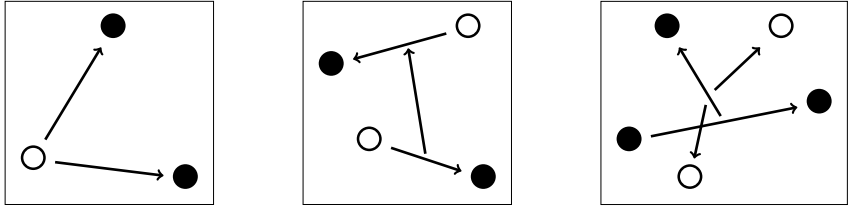}
  \caption{%
    Example translationally invariant coordinates: 
    coordinates of relative vectors within the many-particle system.
    \label{fig:coortrans}
  }
\end{figure}

\subsection{Hamiltonian}
Translationally invariant energies and wave functions are computed 
by writing the kinetic energy operator in TI Cartesian coordinates defined in  
Eqs.~(\ref{eq:ticoord})--(\ref{eq:umx}) (see also Figure~\ref{fig:coortrans})
and subtracting the kinetic energy of the center of mass, 
which results in a translationally invariant Hamiltonian \cite{MaRe12}.
An alternative approach has been proposed in Ref.~\cite{SiMaRe13,MuMaRe17}, which
obviates the need of performing any coordinate transformation and eliminates
any translational contamination in the matrix elements
of the $(\npart+1)$-particle kinetic energy operator, Eq.~(\ref{eq:kin}),
during the integral evaluation.

\subsection{Basis functions}
We approximate the eigenfunction corresponding to 
some spatial $\lambda=(NM_Np)$ and spin $\varsigma=(S_a,M_{S_a},S_b,M_{S_b},\ldots)$ 
quantum numbers as a linear combination of symmetry-adapted basis functions 
\begin{align}
  \Psi^{[\lambda,\varsigma]}
  = 
  \sum_{I=1}^{N_\mr{b}} 
    c_I
    \Phi_{I}^{[\lambda,\varsigma]} .
  \label{eq:lincomb}
\end{align}
The $I$th basis function is a(n) (anti)symmetrized product of spatial and spin
functions for (fermions) bosons  
\begin{align}
  \Phi_{I}^{[\lambda,\varsigma]}(\mx{r},\mx{\sigma}) %
  =
  \hat{\mathcal{A}}\lbrace\phi_{I}^{[\lambda]}(\mx{r}) \chi_I^{[\varsigma]}(\mx{\sigma}) \rbrace
\end{align}
with the (anti)symmetrization operator:
\begin{align}
  \hat{\mathcal{A}} 
  = 
  (N_\mr{perm})^{-1/2}
  \sum_{p=1}^{N_\mr{perm}}
    \varepsilon_p \hat{P}_p ,
\end{align}
where $N_\mr{perm}$ is the total number of possible permutations of the identical particles 
in the system
and 
$\varepsilon_p$ is $-1$ if the permutation operator, $\hat{P}_p$, contains an odd number of
interchanges of identical fermions, otherwise $\varepsilon_p$ is $+1$. 

In order to define spatial basis functions, we first introduce geminal (or pair) functions as
\begin{align}
  \varphi(\mx{r}_1,\mx{r}_2)
  &=
  \exp\left(-\frac{1}{2}\alpha_{12}(\mx{r}_1-\mx{r}_2)^2\right) \\
  &=
  \exp\left(-\frac{1}{2}\mx{r}\tr(\mx{A}\otimes\mx{I}_3)\mx{r}\right) , 
  \label{eq:mxgem}
\end{align}
with
\begin{align}
  \mx{r}
  =
  \left(%
  \begin{array}{@{}c@{}}
    \mx{r}_1 \\       
    \mx{r}_2 \\
  \end{array}
  \right)\in\mathbb{R}^{6}
  \quad
  \text{and}
  \quad
  \mx{A}
  =
  \left(%
  \begin{array}{@{}cc@{}}
    \mx{A}_{11} & \mx{A}_{12} \\ 
    \mx{A}_{21} & \mx{A}_{22} \\ 
  \end{array}
  \right)
  =
  \left(%
  \begin{array}{@{}cc@{}}
     \alpha_{12} & -\alpha_{12} \\ 
    -\alpha_{12} &  \alpha_{12} \\ 
  \end{array}
  \right)\in\mathbb{R}^{2\times 2}.
\end{align}
The geminal functions are generalized to 
$(\npart+1)$-particle explicitly correlated Gaussian functions (ECGs) as:
\begin{align}
  \phi(\mx{r};\mx{A})
  &=
  \prod_{i=1}^{\npart+1}
  \prod_{j>i}^{\npart+1}
  \exp\left(-\frac{1}{2}\alpha_{ij}(\mx{r}_i-\mx{r}_j)^2\right) \\
  &=
  \exp\left(%
    -\frac{1}{2} \mx{r}\tr(\mx{A}\otimes \mx{I}_3)\mx{r}
  \right) 
  \label{eq:ecg}
\end{align}
with
\begin{align}
  \mx{r}
  =
  \left(%
  \begin{array}{@{}c@{}}
    \mx{r}_1 \\
    \ldots \\
    \mx{r}_{\npart+1} \\
  \end{array}
  \right)\in\mathbb{R}^{3(\npart+1)}
\end{align}
and
\begin{align}
  A_{ij}
  =
  -\alpha_{ij}(1-\delta_{ij})
  + 
  \left(\sum_{k=1,k\neq i}^{\npart+1} \alpha_{ij}\right)\delta_{ij} 
  \quad 
  (i,j=1,\ldots,\npart+1).
\end{align}
The matrix form, Eqs.~(\ref{eq:mxgem}) and (\ref{eq:ecg}), 
of the functions makes it apparent that the functions
have a general mathematical form for $(\npart+1)$ particles.
It has been observed \cite{MaRe12} that 
in molecular applications with multiple heavy particles (nuclei), 
these basis functions are inefficient when
sub-spectroscopic accuracy is sought for.
In a first attempt to describe atomic nuclei more efficiently, we may introduce
ECGs with shifted centers, so called ``floating ECGs'':
\begin{align}
  \phi(\mx{r};\mx{A},\mx{\mathcal{R}})
  =
  \exp\left(%
    -\frac{1}{2} %
    (\mx{r}-\mx{\mathcal{R}})\tr %
    (\mx{A}\otimes \mx{I}_3) %
    (\mx{r}-\mx{\mathcal{R}})
  \right) ,
  \label{eq:fecg}
\end{align}
where $\mathcal{R}$ can be treated as a fixed or a variational parameter
and thereby, it should provide a more efficient description for the atomic nuclei displaced 
from the origin (Section~\ref{ch:molstruct}).
It turns out however that the convergence of a molecular computation 
is even worse with the floating ECGs, Eq.~(\ref{eq:fecg}),
than with origin-centered ECGs, Eq.~(\ref{eq:ecg}).
This behaviour is due to the fact that
$\phi(\mx{r};\mx{A},\mx{\mathcal{R}})$ with an arbitrary $\mx{\mathcal{R}}\neq 0$ is 
not an eigenfunction of neither the total angular momentum operators, 
$\hat{N}^2$ and $\hat{N}_z$, nor the parity.

In order to obtain very accurate numerical results for a molecular system,
it is necessary to describe displaced atomic nuclei efficiently, 
and at the same time, account for the spatial
symmetries of the system. In principle, it would be possible to project the
floating ECGs, Eq.~(\ref{eq:fecg}), onto the irreps of the O(3) group 
by numerical projection. 
As an alternative, we use explicitly correlated Gaussians 
in conjunction with the global vector representation (GVR-ECG) \cite{VaSu95,SuUsVa98,SuVaBook98}:
\begin{align}
  \phi^{[\lambda]}(\mx{r};\mx{A},\mx{u},K)
  &= 
  \frac{1}{B_{KN}}
  \int\dd\hat{\mx{e}}\ Y_N^{M_N}(\hat{\mx{e}})
  \left\lbrace
    \partial_a^{(2K+N)}
    g(\mx{r};\mx{A},a\mx{u}\otimes\mx{e})
  \right\rbrace_{a=0,|\mx{e}|=1},
  \label{eq:gvr2} \\
  &= 
  |\mx{v}|^{2K+N} Y_{N}^{M_N}(\hat{\mx{v}}) %
  \exp\left(-\frac{1}{2}\mx{r}\tr(\mx{A}\otimes\mx{I}_3)\mx{r}\right), 
  \label{eq:gvr}
\end{align}
which corresponds to an analytic projection of the generator function
\begin{align}
  g(\mx{r};\mx{A},a\mx{u}\otimes\mx{e})
  =
  \exp\left(%
    -\frac{1}{2}\mx{r}\tr
    (\mx{A}\otimes\mx{I}_3)
    \mx{r}
    +a(\mx{u}\otimes\mx{e})\tr\mx{r}
  \right)
\end{align}
with the global vector
\begin{align}
  \mx{v}
  =
  (\mx{u}\otimes \mx{e})\tr\mx{r}
  =
  \sum_{i=1}^{\npart+1}u_i\mx{r}_i .
  \label{eq:globalvec}
\end{align}

Further notation used in Eqs.~(\ref{eq:gvr})--(\ref{eq:gvr2}):
$\partial_a^{(2K+N)}=\partial^{(2K+N)}/\partial a^{(2K+N)}$,  
$Y_{N}^{M_N}(\hat{\mx{e}})$ is the spherical harmonic function of degree $N$ and order $M_N$,
$\hat{\mx{e}}=(\theta,\phi)$ collects the polar angles characterizing the orientation of 
the unit vector $\mx{e}$, and
\begin{align}
  B_{KN} 
  = 
  \frac{%
    4\pi\ (2K+N)! (K+N+1)\ 2^{N+1}
   }{%
    K!\ (2K+2N+2)!
   }
\end{align}
with $K$ and $N\in\mathbb{N}_0.$
According to Ref.~\cite{SuUsVa98} the application of
Eq.~(\ref{eq:gvr}) in a variational procedure is equivalent to using a 
basis set constructed by a hierarchical coupling of the subsystems angular 
momenta to a total angular momentum state with $(N,M_N)$.
Thereby, $\phi^{[\lambda]}(\mx{r};\mx{A},\mx{u},K)$ 
is eigenfunction of $\hat{N}^2$ and $\hat{N}_z$, as well as the space-inversion operator.

Relation of the generator function to a floating ECG becomes apparent by writing
\begin{align}
  &\exp\left(%
    -\frac{1}{2}
    (\mx{r}-\mx{\mathcal{R}})\tr
    (\mx{A}\otimes\mx{I}_3)
    (\mx{r}-\mx{\mathcal{R}})
  \right) 
  \nonumber \\
  =
  &\exp\left(%
    -\frac{1}{2}\mx{\mathcal{R}}\tr(\mx{A}\otimes\mx{I}_3)\mx{\mathcal{R}}
  \right)
  \exp\left(%
    -\frac{1}{2}\mx{r}\tr(\mx{A}\otimes\mx{I}_3)\mx{r}
    + \mx{\mathcal{R}}\tr(\mx{A}\otimes\mx{I}_3)\mx{r}
  \right).
\end{align}
Alternatively, the form of $\phi^{[\lambda]}(\mx{r};\mx{A},\mx{u},K)$ in Eq.~(\ref{eq:gvr})
contains a polynomial prefactor, 
which describes efficiently particles displaced from the center of mass.

We use the $\phi^{[\lambda]}(\mx{r};\mx{A},\mx{u},K)$ as spatial basis function
and optimize the (non-linear) parameters
$\alpha_{ij}\in\mathbb{R}$ ($\mx{A}$), $u_i\in\mathbb{R}$ ($\mx{u}$), $K\in\mathbb{N}_0$ 
variationally.
The transformation properties of these functions 
under TI coordinate transformations are equivalent to a simple
transformation of the parameter vectors ($\mx{u}$ in the global vector) 
and matrices ($\mx{A}$ in the exponent) summarized in the following equations:
\begin{align}
  &|v|^{2K+N}Y_{N}^{M_N}(\hat{\mx{v}})
  \exp\left(-\frac{1}{2}\mx{r}\tr(\mx{A}\otimes\mx{I}_3)\mx{r}\right) \\
  =\ 
  &|v|^{2K+N}Y_{N}^{M_N}(\hat{\mx{v}})
  \exp\left(-\frac{1}{2}\mx{x}\tr(\mx{\mathcal{A}}^{(x)}\otimes\mx{I}_3)\mx{x}\right) \\
  =\ 
  &|v|^{2K+N}Y_{N}^{M_N}(\hat{\mx{v}})
  \exp\left(-\frac{1}{2}\mx{y}\tr(\mx{\mathcal{A}}^{(y)}\otimes\mx{I}_3)\mx{y}\right), 
\end{align}
with
\begin{align}
  \mx{A}^{(x)}=\mx{U}^{-\mr{T}}\mx{A}\mx{U}^{-1} 
  \quad&\Leftrightarrow\quad
  \mx{A}=\mx{U}^{\mr{T}}\mx{A}^{(x)}\mx{U} 
  \quad\text{and}\quad
  \mx{A}^{(x)}
  =
  \left(%
    \begin{array}{@{}cc@{}}
      \mx{\mathcal{A}}^{(x)} & 0 \\
      0 & c_A \\
    \end{array}
  \right)
  \\
  \mx{A}^{(y)}=\mx{V}^{-\mr{T}}\mx{A}\mx{V}^{-1} 
  \quad&\Leftrightarrow\quad
  \mx{A}=\mx{V}^{\mr{T}}\mx{A}^{(y)}\mx{V}
  \quad\text{and}\quad
  \mx{A}^{(y)}
  =
  \left(%
    \begin{array}{@{}cc@{}}
      \mx{\mathcal{A}}^{(y)} & 0 \\
      0 & c_A \\
    \end{array}
  \right)  
\end{align}
and 
\begin{align}
  \mx{A}^{(y)} = (\mx{U}\mx{V}^{-1})\tr\mx{A}^{(x)}\mx{U}\mx{V}^{-1}
  \quad\Leftrightarrow\quad
  \mx{A}^{(x)} = (\mx{V}\mx{U}^{-1})\tr\mx{A}^{(y)}\mx{V}\mx{U}^{-1} .
\end{align}
and the global-vector coefficients transform as
\begin{align}
  \mx{u}
  = \mx{U}\tr\mx{u}^{(x)}
  = \mx{V}\tr\mx{u}^{(y)}
  \quad\text{and}\quad
  \mx{u}^{(x)}
  = (\mx{U}\mx{V}^{-1})\tr \mx{u}^{(y)} .
\end{align}
These simple transformation properties 
are exploited during the course of the analytic calculation of the matrix elements 
of the kinetic and potential energy operators 
\cite{MaHuMuRe11a,MaHuMuRe11b,MaRe12,Ma13,SiMaRe13,SiMaRe14}.

\subsection{Evaluation of the matrix elements}
Matrix elements of an $\hat{O}$ operator---\emph{e.g.,}
identity, kinetic or potential energy operators,
$\hat{O}=\hat{I},\hat{T}$ or $\hat{V}$, respectively---with 
the (anti)symmetrized products of 
the spin and spatial functions, Eqs.~(\ref{eq:gvr2})--(\ref{eq:gvr}),
are obtained by analytic expressions. 
In what follows, we summarize the main steps of the calculations 
(further details can be found in Refs.~\cite{SuVaBook98,MaRe12,JoVa16}).
\begin{align}
  O_{IJ}^{[\lambda,\varsigma]}
  &=
  \langle %
    \Phi_{I}^{[\lambda,\varsigma]} %
    |\hat{O}| %
    \Phi_{J}^{[\lambda,\varsigma]}  %
  \rangle_{\mr{r},\sigma} \nonumber \\
  &=
  \langle %
    \hat{\mathcal{A}}\lbrace\phi_{I}^{[\lambda]}\chi_I^{[\varsigma]}\rbrace %
    |\hat{O}| %
    \hat{\mathcal{A}}\lbrace\phi_{J}^{[\lambda]}\chi_J^{[\varsigma]}\rbrace %
  \rangle_{\mr{r},\sigma} \nonumber \\
  &=
  \sum_{p=1}^{N_\mr{perm}}
  \varepsilon_p
  \langle %
    \phi_{I}^{[\lambda]}\chi_I^{[\varsigma]} %
    |\hat{O}| %
    \hat{P}_p\lbrace\phi_{J}^{[\lambda]}\chi_J^{[\varsigma]}\rbrace %
  \rangle_{\mr{r},\sigma} \nonumber \\
  &=
  \sum_{p=1}^{N_\mr{perm}}
  \varepsilon_p
  \langle %
    \phi_{I}^{[\lambda]} %
    |\hat{O}| %
    \hat{P}_p\phi_{J}^{[\lambda]} %
  \rangle_{\mr{r}} %
  \langle %
    \chi_I^{[\varsigma]} %
    |\hat{O}| %
    \hat{P}_p\chi_J^{[\varsigma]} %
  \rangle_{\sigma} \nonumber \\
  &=
  \sum_{p=1}^{N_\mr{perm}}
    c_{IJ_p}^{[\varsigma]} O_{IJ_p}^{[\lambda]}
\end{align}
with the terms
\begin{align}
  c_{IJ_p}^{[\varsigma]}
  =
  \varepsilon_p
  \langle %
    \chi_I^{[\varsigma]} %
    |\hat{O}| %
    \hat{P}_p\chi_J^{[\varsigma]} %
  \rangle_{\sigma} 
  \quad\text{and}\quad
  O_{IJ_p}^{[\lambda]}
  =
  \langle %
    \phi_{I}^{[\lambda]} %
    |\hat{O}| %
    \phi_{J_p}^{[\lambda]} %
  \rangle_{\mr{r}} , %
\end{align}
which are separate integrals of $\hat{O}$ with the spin and the spatial functions, respectively.
The $c_{IJ_p}^{[\varsigma]}$ term can be obtained by simple algebra (see for example
Ref.~\cite{MaRe12}). The $O_{IJ_p}^{[\lambda]}$ term contains multidimensional integrals of 
the spatial functions, Eq.~(\ref{eq:gvr2})--(\ref{eq:gvr}), for which analytic expressions 
are obtained by working out the formal operations in three steps. \\
\noindent\emph{Step~1:} evaluation of the integral with the generator function:
  \begin{align}
    I_{O,1}(\mx{s},\mx{s}')
    =
    \langle %
      g(\mx{r};\mx{A},\mx{s})
      |\hat{O}|
      g(\mx{r};\mx{A}',\mx{s}')
    \rangle_{\mx{r}} 
  \end{align}
\emph{Step~2:} expansion of the angular pre-factors:
  \begin{align}
  I_{O,2}(\mx{e},\mx{e}')
  =
  \lbrace %
    \partial_a^{2K+N} %
    \partial_{a'}^{2K'+N} %
      I_{O,1}(a\mx{u}\otimes\mx{e},a'\mx{u}'\otimes\mx{e}')
  \rbrace_{%
  {\tiny%
  \begin{array}{@{}l@{}}
    \tiny a=a'=0 \\
    \tiny |\mx{e}|=|\mx{e}'|=1 \\
  \end{array}
  }
  } 
  \end{align}
\emph{Step~3:} evaluation of the angular integrals:
  \begin{align}
  O^{[\lambda]}
  =
  \displaystyle
  \frac{1}{B_{KN}B_{K'N}}
  \int\dd\hat{\mx{e}}\int\dd\hat{\mx{e}}'\ 
    (Y_N^{M_N}(\hat{\mx{e}}))^\ast
    Y_N^{M_N}(\hat{\mx{e}'})\ 
    I_{O,2}(\mx{e},\mx{e}')
  \end{align}
  
The resulting expressions \cite{SuVaBook98,MaRe12,JoVa16} are completely general 
for basis function with any $N$ total angular momentum 
quantum number and natural parity, $(-1)^N$ (similar functions and working formulae 
with unnatural parity, $(-1)^{N+1}$, were introduced in Ref.~\cite{SuHoOrAr08}). 
Details of the computer implementation of the final expressions are critical 
in order to ensure the numerical stability and efficiency of molecular applications
in which $N$ and $K$ are larger than about 5.
In particular, we had to introduce the so-called ``quasi-normalization'' of the basis functions and 
pre-compute and tabulate certain coefficients with infinite-precision arithmetics 
to be able to evaluate the expressions 
in double precision arithmetics 
(this implementation was tested up to ca. $2K=40$ and $N=5-10$) \cite{MaRe12}.

\subsection{Computation of bound states}

\subsubsection{Direct diagonalization}
The $c_I$ linear combination coefficients in Eq.~(\ref{eq:lincomb}) are obtained 
by solving the generalized eigenvalue problem:
  \begin{align}
    \mx{H} \mx{c}_I = E_I \mx{S}\mx{c}_I .
    \label{eq:bound1}
  \end{align}
The eigenvalues and eigenvectors are computed by replacing 
Eq.~(\ref{eq:bound1}) with the symmetric eigenvalue equation
  \begin{align}
    \mx{H}' \mx{c}'_I = E_I \mx{c}'_I
    \label{eq:bound2}
  \end{align}
by using L\"owdin's procedure \cite{MayerBook03}
  \begin{align}
    \mx{H}'=\mx{T}'+\mx{V}'\ 
    \label{eq:hmx}    
  \end{align}
with
  \begin{align}
    \mx{T}'
    =
    \mx{S}^{-1/2}\mx{T}\mx{S}^{-1/2}
    \quad\quad
    \text{and}
    \quad\quad
    \mx{V}'
    =
    \mx{S}^{-1/2}\mx{V}\mx{S}^{-1/2}.
    \label{eq:tvmx}    
  \end{align}

\subsubsection{Non-linear parameterization strategy}
  \paragraph{Parameter selection}
  The nonlinear parameters for each basis function are 
  selected and optimized based on the variational principle applicable for 
  the ground and a finite number of excited states 
  (p. 27--29 of Ref.~\cite{SuVaBook98}), 
  which translates in practice to the simple rule: 
  the lower the energy, the better the parameter set.
  The parameter selection is carried out using 
  the stochastic variational method \cite{SuVaBook98}, 
  in which new basis functions are generated one by one. 
  Trial values for the parameters of the spatial basis functions, Eq.~(\ref{eq:gvr2}), 
  $K$, $u_i$, $\ln\alpha_{ij}$, are drawn from 
  discrete uniform, continuous uniform, and normal distributions, respectively. 
  The optimal parameters of each distribution
  are estimated from short exploratory computations. 
  Due to the one-by-one generation of the basis functions, 
  the updated eigenvalues can be evaluated very efficiently 
  \cite{SuVaBook98} 
  using the known eigenvalues and eigenvectors corresponding to 
  the old basis set, and this allows for a rapid assessment of a trial parameter set.
  
  \paragraph{Refinement}
  The refinement of the basis-function parameters generated by 
  the stochastic variational method is necessary if very accurate solutions are required.
  Similarly to the enlargement of the basis set, 
  the basis functions are refined one after the other using 
  the fast rank-1 eigenvalue update algorithm, 
  which is used also for the selection of a new basis function from 
  a set of randomly generated trials. 
  Refined parameters are found by using the Powell method \cite{Po04} 
  started from the originally selected parameters for each basis function. 
  The random-walk refinement can be used to adjust the $K$ integer value 
  (for which the Powell method is not applicable), 
  however in practice it is usually sufficient 
  to generate $K$ from a discrete uniform distribution spread over 
  a pre-optimized interval and to refine only the continuous variables, 
  $u_i$ and $\alpha_{ij}$ by the Powell method. During the course of and at the end
  of the enlargement of the basis set, 
  every basis function is refined in repeated cycles.

\subsection{Computation of resonance states}
  \subsubsection{Stabilization technique}
    The stabilization of eigenvalues of the real eigenvalue equation, Eq.~(\ref{eq:bound1}), 
    is monitored with respect to the size of the basis set \cite{UsSu02,SuUs04}. 
    This simple application of the stabilization
    method \cite{HaTa70,TaHa76,MaTaRyMo94,MuYaBu94} allowed us to estimate 
    the energy of long-lived resonances \cite{Ma13}. In order to gain access to 
    the lifetimes (and in general, shorter-lived resonance positions and widths), 
    it is necessary to estimate the box size corresponding to the increasing number of 
    basis functions, which is a non-trivial task with ECG functions.

  \subsubsection{Complex-coordinate-rotation method}
    The application of the complex-coordinate-rotation method \cite{KuKrHo88} requires 
    the complex scaling of the coordinates according to the 
    $r \rightarrow r\mr{e}^{\mr{i}\theta}$ replacement. The scaling rule is rather simple 
    for both the kinetic energy and the Coulomb potential energy operators, 
    and thus the Hamiltonian is scaled according to 
    \begin{align}
          \hat{H} 
          =
          \hat{T} + \hat{V} 
          \quad\rightarrow\quad
          \hat{H}(\theta)
          =
          \mr{e}^{-2\mr{i}\theta} \hat{T}
          +
          \mr{e}^{-\mr{i}\theta} \hat{V} .
    \end{align}
    The corresponding matrix equation is written as
    \begin{align}
      \mx{\tilde H}(\theta) \mx{\tilde c}_i(\theta)
      = 
      \mathcal{E}_i(\theta) \mx{S}\mx{\tilde c}_i(\theta) ,
      \label{eq:reseq1}
    \end{align}
    which, similarly to its real analogue, Eq.~(\ref{eq:bound1}), is transformed to
    \begin{align}
      \mx{\tilde H}'(\theta) \mx{\tilde c}'_i(\theta) 
      = 
      \mathcal{E}_i(\theta) \mx{\tilde c}'_i(\theta)
      \label{eq:reseq2}
    \end{align}
    with
    \begin{align}
      \mx{\tilde H}'(\theta) 
      &= 
      \mr{e}^{-2\mr{i}\theta} \mx{S}^{-1/2}\mx{T}\mx{S}^{-1/2}
      +
      \mr{e}^{-\mr{i}\theta} \mx{S}^{-1/2}\mx{V}\mx{S}^{-1/2} \nonumber \\
      &= 
      \cos(2\theta) \mx{T}' + \cos(\theta)\mx{V}'
      -\mr{i}(\sin(2\theta)\mx{T}' + \sin(\theta)\mx{V}') .
    \end{align}
    The complex symmetric eigenproblem, Eq.~(\ref{eq:reseq2}), is solved using 
    LAPACK library routines \cite{lapack}, and the stabilization point, 
    $\mathcal{E}=(E,-\Gamma/2)$ with the $E$ energy and $\Gamma$ width,
    on the complex energy plane is identified visually.
    
    Although the complex analogue of the real variational principle \cite{KuKrHo88} states that 
    the exact solution is a stationary point on the complex plane with respect to the 
    variational parameters and the scaling angle, 
    there is not any practical algorithm for using this principle to optimize the basis set 
    and to systematically improve the resonance parameters.
    The convergence of the resonance parameters is confirmed by 
    achieving reasonable agreement within a series of computations 
    with a varying number of basis functions and 
    parameterization (see also Sec.~\ref{sec:numex}).

  \subsubsection{Parameterization strategy}
    Due to the lack of any practical approach relying on the complex variational principle 
    to select and optimize the non-linear parameters of the basis functions, 
    we relied on the random generation of the parameters from some broad parameter intervals. 
    In addition, we have devised
    a parameter-transfer approach \cite{Ma13}, in which a parameter set optimized 
    based on the real variational principle for bound states
    with one set of input parameters is transferred to a computation 
    with other input parameters (\emph{e.g.,} different quantum numbers). 
    Note that the spatial symmetries of a basis function 
    are determined by the quantum numbers, Eq.~(\ref{eq:gvr}), 
    and in this sense, the parameters $K$, $u_i$, and $\mx{A}$, are transferable.

%
%
\subsection{Variational results \label{sec:numex}}

\begin{figure}
  \includegraphics[scale=1.]{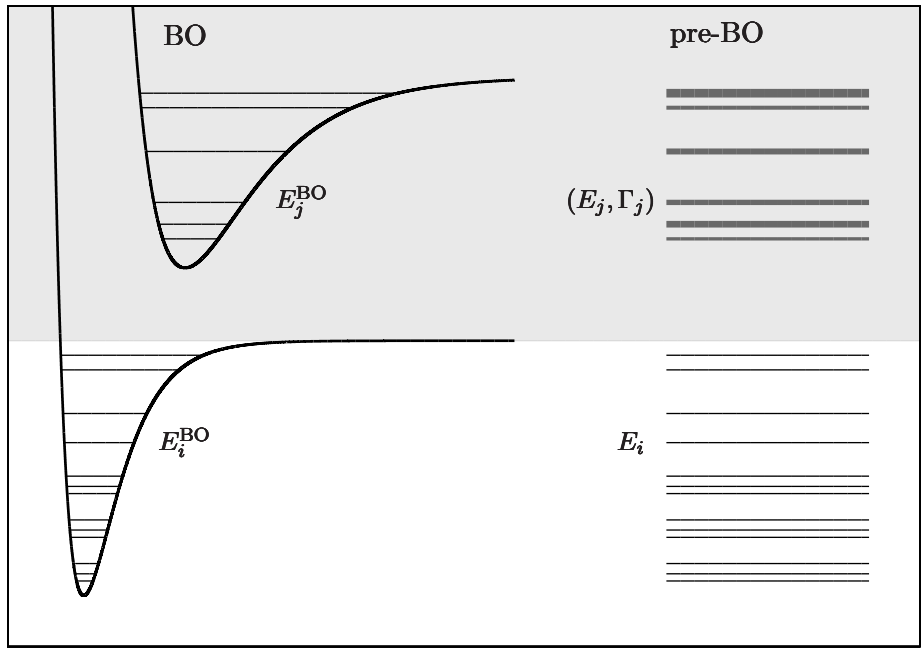}
  \caption{%
    The ladder structure of the pre-Born--Oppenheimer (pre-BO) energy levels (right). 
    The left of the figure visualizes the rovibrational states corresponding to 
    their respective potential energy surfaces in the Born--Oppenheimer (BO) approximation.
    While in the BO picture, the rovibrational states corresponding to 
    the excited electronic state are bound states, 
    they appear as resonances in the full pre-BO treatment.    
  [Reprinted with permission from E. Mátyus, J. Phys. Chem. A 117, 7195 (2013).
  Copyright 2013 American Chemical Society.]
    \label{fig:res}    
  }
\end{figure}

Quantitative comparison of precision experiments and computations
is possible if 
extremely accurate non-relativistic results are available and 
they are corrected also
for relativistic and quantum electrodynamics (QED) effects.
Such corrections have been computed for bound states of few-particle systems within
a perturbative scheme, which started from 
a very accurate Born--Oppenheimer solution \cite{PiJe09}.
As an alternative route, efforts have been devoted to the  
a direct variational solution of the Dirac equation \cite{SiMaRe15,Simmen15}.
As to an intermediate approach, expectation values of the Breit--Pauli Hamiltonian 
were computed with the many-particle wave function, \emph{i.e.,}
without evoking the BO approximation, to obtain relativistic corrections 
with the full many-particle wave function \cite{StAd13,StJuAd17,KoHiKa06}. 
In the spirit of the second direction, 
the present work focuses on the computational methodology 
of very accurate non-relativistic energies and wave functions, 
which provides the starting point for a forthcoming computation of 
relativistic and QED corrections.
 
Within the non-relativistic regime, the determination of not only the ground 
but also the excited states with all possible combinations of 
the quantum numbers is a challenging task.
What makes it particularly challenging is the fact that
rovibrational states corresponding to excited electronic states, 
which appear as bound states within the BO approximation, 
are rigorously obtained as resonances fully coupled 
to the dissociation continuum of the lower-lying electronic states (Figure~\ref{fig:res}).
This makes the computation and a systematic improvement 
of excited rovibronic states (with various non-relativistic quantum numbers) a 
highly challenging task. Nevertheless, once it is successfully solved, 
not only the energy position but also 
the predissociative lifetime is obtained, potentially 
from a full pre-BO computation.
The following paragraphs review variational results obtained 
in a series of computations \cite{MaRe12,Ma13} motivated by these ideas.

\vspace{0.5cm}
\paragraph{Bound and resonances states of the positronium molecule, 
Ps$_2=\lbrace\text{e}^+,\text{e}^+,\text{e}^-,\text{e}^-\rbrace$}
Computation of positronium complexes are extremely challenging 
for traditional quantum chemistry methods,
because of the presence of positively charged light particles. 
At the same time, positronium complexes are excellent test systems for pre-BO methodological 
developments \cite{MaRe12,Ma13}. The present GVR-ECG basis set
has turned out to be particularly well-suited for positronium systems, 
which is explained qualitatively by their diffuse, delocalized internal structure 
in comparison with the localized atomic nuclei in molecular systems. 
Tightly converged energy levels were computed with
basis functions including only low-order polynomial prefactors in Ref~\cite{Ma13}, 
(Thanks to the low-order polynomials in the basis functions, Eq.~(\ref{eq:gvr2})--(\ref{eq:gvr}). 
the results were obtained with only modest computational cost and this fact makes 
the positronium an excellent test for testing and developing the pre-BO method.)
The basis function parameters were selected by minimizing 
the energy of the lowest-lying state. The resulting basis set 
was well-suited for not only the lowest-energy bound state 
but also for a few low-energy resonance states.

The obtained bound-state energies and resonance parameters 
were in excellent agreement or improved upon 
the best results available in the literature (see Table~2 of Ref.~\cite{Ma13}). 
We may think that the computed resonance parameters were more accurate 
than earlier literature data, because the energy of nearby-lying bound states
was improved for which the (real) variational principle allows us to make
a clear-cut assessment.

\vspace{0.5cm}
\paragraph{Bound and resonances states of the hydrogen molecule, H$_2=\lbrace\text{p}^+,\text{p}^+,\text{e}^-,\text{e}^-\rbrace$}
The first variational computations with
explicitly correlated Gaussian function and $N>0$ angular momentum quantum numbers
carried out for the H$_2$ molecule as a four-particle system 
were reported in Refs.~\cite{MaRe12,Ma13}.
Both the ground and certain excited electronic states were considered.
We note that exceedingly accurate pure vibrational states of the ground electronic
state were computed earlier by the Adamowicz group \cite{BuAd03}. 
Furthermore, very accurate rovibrational
states corresponding to the ground electronic states were available from the non-adiabatic
perturbation theory computations performed by Pachucki and Komasa \cite{PaKo09}.

Besides the ground electronic state, excited electronic states became accessible 
in independent computations by choosing different combinations of 
the non-relativistic quantum numbers in Ref.~\cite{Ma13}, where 
four different blocks with natural parity were computed:
\begin{itemize}
  \item[ ]
    ``$\X1Sgp$ block'': $N\geq 0,\ p=(-1)^N,\ \Sp=(1-p)/2,\ \Se=0$;
  \item[ ]
    ``$\B1Sup$ block'': $N\geq 0,\ p=(-1)^N,\ \Sp=(1+p)/2,\ \Se=0$; 
  \item[ ]
    ``$\a3Sgp$ block'': $N\geq 0,\ p=(-1)^N,\ \Sp=(1-p)/2,\ \Se=1$;
  \item[ ]
    ``$\b3Sup$ block'': $N\geq 0,\ p=(-1)^N,\ \Sp=(1+p)/2,\ \Se=1$.
\end{itemize}

The computations resulted in improved energies for some of the rotational states corresponding 
to electronically excited states (see Table~3 in Ref.~\cite{Ma13}).
For the lowest rotational states of the $\B1Sup$ block the newly computed energies were 
lower than those of Ref.~\cite{WoOrSt06} by ca.~$0.8~\mu$\Eh. Furthermore,
the computed energies improved upon the first and the second rotational states
of the $\a3Sgp$ block by a few tens of n\Eh\ in comparison with the best earlier 
prediction \cite{Wo07}.

\begin{figure}
  \begin{center}
    \includegraphics[scale=1.]{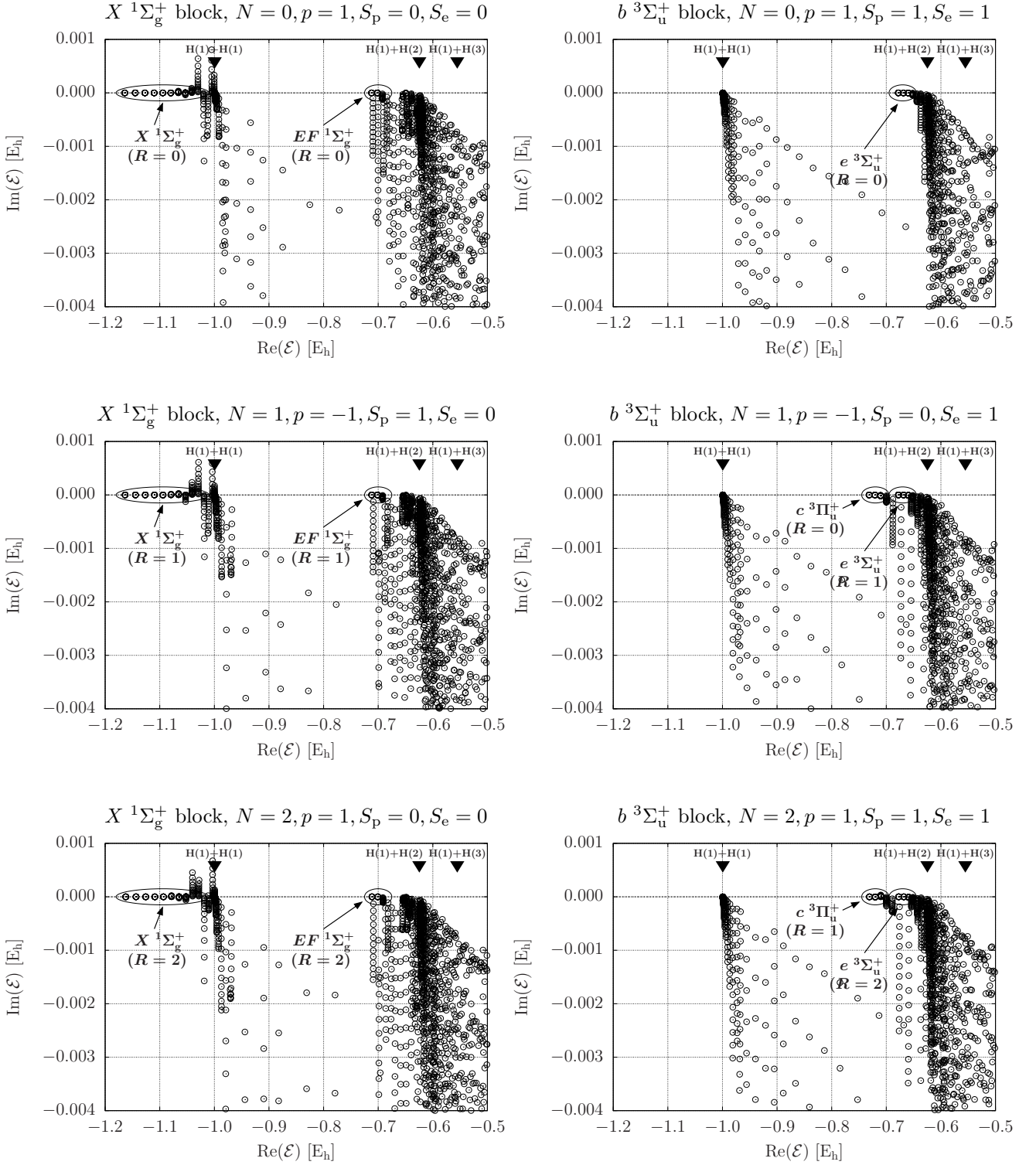} \\
  \end{center}
  \caption{%
    Part of the spectrum of the complex scaled Hamiltonian, 
    $\mathcal{H}(\theta)$ with $\theta\in[0.005,0.065]$
    for the $X\ ^1\Sigma_\mr{g}^+$ block $[p=(-1)^N,S_\mr{p}=(1-p)/2,S_\mr{e}=0]$ 
    and 
    for the $b\ ^3\Sigma_\mr{u}^+$ block $[p=(-1)^N,S_\mr{p}=(1+p)/2,S_\mr{e}=1]$
    with $N=0,1,$ and $2$ total spatial angular momentum quantum numbers.
    The black triangles indicate the threshold energy of the dissociation continua corresponding to
    H(1)+H(1), H(1)+H(2), and H(1)+H(3).
  [Reprinted with permission from E. Mátyus, J. Phys. Chem. A 117, 7195 (2013).
  Copyright 2013 American Chemical Society.]
    \label{fig:resonH2}
  }
\end{figure}

In comparison with the positronium molecule, the basis-set parameterization for 
the hydrogen molecule has turned out to be computationally far more demanding for 
the bound states
and a really challenging task for the resonances. 
As to rovibrational (rovibronic) states corresponding to higher excited electronic states, 
they can be computed within the pre-BO framework as resonances embedded 
in the continuum of the lowest-energy electronic state of their respective symmetry block 
(Figure~\ref{fig:resonH2}). 
At present, there is not any practical approach for the optimization of 
basis functions for resonance states.
Instead, we followed a practical strategy to gain access to some resonance states
by compiling a giant parameter set 
from all parameters obtained in bound-state optimizations with 
various combinations of the non-relativistic quantum numbers, 
and performed a search for resonance states using this large set 
(parameter-transfer approach).
The stabilization of certain points on the complex plane with respect to 
the scaling angle are visualized in Figure~\ref{fig:resonH2} for the 
$\X1Sgp$ and $\b3Sup$ blocks with $N=0,1,$ and 2 angular momentum quantum numbers. 

The $\X1Sgp$ block starts with the bound (ro)vibrational states corresponding to 
the ground electronic state, $\X1Sgp$, which are along the real axis up 
to the first dissociation threshold, $\text{H}(1)+\text{H}(1)$, 
indicated with a black arrow in each subfigure. Before the start of the second dissociation limit, 
$\text{H}(1)+\text{H}(2)$, we identify (ro)vibrational states corresponding to the 
$\EF1Sgp$ electronic state.

As to the $\b3Sup$ block, the figure shows that it starts with the 
the first dissociation channel, $\text{H}(1)+\text{H}(1)$, and does not support
any bound state (in agreement with our knowledge from BO and post-BO results). 
Before the $\text{H}(1)+\text{H}(2)$ channel opens, 
we observe a series of vibrational states for $N=0$, which can be assigned  
(based on their energies) to the $\e3Sup$ electronic state.
These states are located very close to the real axis, 
which indicates that they are long-lived resonances. 
It is interesting to note the appearance of a set of lower-energy states for $N>0$.
This set of states can be assigned to the vibrational ($R=0$ rotational angular momentum) 
and rovibrational ($R=1$) states corresponding to the $\c3Pup$ electronic state 
(with $L=1$ orbital angular momentum) for $N=1$ and 2, respectively.
This example highlights the coupling of the 
electronic orbital ($\hat{\vec{L}}$) and rotational angular ($\hat{\vec{R}}$) momenta 
to the total angular momentum ($\hat{\vec{N}}$), 
which is automatically included in our pre-BO approach. 
We note that the electronic, $L$, and rotational, $R$, angular momentum quantum numbers are 
non-exact quantum numbers in the full many-particle quantum treatment, 
but they are useful labels to describe properties of a state with 
some $N$ total angular momentum quantum number.
 
Ref.~\cite{Ma13} gives a detailed account of the numerical results in comparison with
the best available results in the literature: 
accurate adiabatic computations had been performed by Ko{\l}os and Rychlewski 
for the $\e3Sup$ state \cite{KoRy90},
and accurate BO calculations are available for the $\c3Pup$ state from the same authors \cite{KoRy77}. 
Ref.~\cite{Ma13} reported the first computational results of rotationally excited levels
corresponding to the $\c3Pup$ electronic state, 
which can be obtained only by accounting for the coupling of rotational and 
electronic angular momenta (automatically included in our method). 

It is also important to note that the results reviewed in the previous paragraphs
provided accurate estimates for the energies. 
In order to pinpoint the widths and the related lifetimes, 
it will be necessary to optimize and/or enlarge the basis (and parameter) set, 
for which the development of a systematically improvable variational approach 
is required for resonances.

%
%
\clearpage
\section{Molecular structure from quantum mechanics \label{ch:molstruct}}
\noindent%
If the BO approximation is not introduced the non-relativistic limit can be, in principle, 
approached arbitrarily close, and when relativistic and QED corrections are also included, 
computations come close or even challenge precision measurements.
It is important to note however that the present-day theoretical foundations 
for the structure of molecules relies on the BO approximation:
the molecular structure is identified with the equilibrium structure, which is defined as
a local minimum of the potential energy surface. 
Interestingly, there is not available any rigorous and practical definition of 
the molecular structure independent of the BO approximation \footnote{%
For example, the IUPAC's Compendium of Chemical Terminology (``Gold Book'') 
defines the \emph{equilibrium geometry} in terms of a potential energy surface, 
but we do not find anything beyond this apart from the definition of 
the primary, secondary, etc. structures of macromolecules. 
Interestingly, \emph{molecular shape} is defined in the Compendium.}.

In relation to the separation of the motion of the electrons and the atomic nuclei, which is
commonplace in quantum chemistry, 
Hans Primas pointed out in his book \cite{Primas81}:
\emph{%
   ``We describe the six degrees of freedom of the ground state of the helium atom
    (considered as 3-particle problem with the center-of-mass motion separated) 
    as a problem of two interacting particles in an external Coulomb potential.
    However, in the case of the molecule H$_2^+$
    we discuss the very same type of differential equation in an entirely different way,
    and split the 6 degrees of freedom into 
    1 vibrational mode, 2 rotational modes, and 3 electronic type degrees of freedom.
    This qualitatively different description does by
    no means follow from a purely mathematical discussion.''
}
\begin{figure}
  \begin{center}
    \includegraphics[scale=1.]{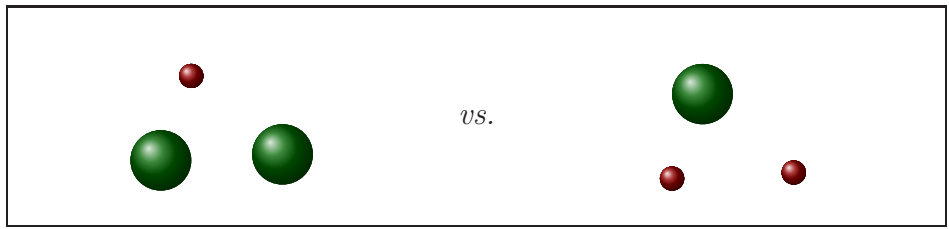}
  \end{center}
  \caption{%
    For the three-particle He atom and for the three-particle H$_2^+$ molecular ion
    in the standard quantum chemistry approach
    \emph{``we discuss the very same type of differential equation in an entirely different way''}
    \cite{Primas81}.
    \label{fig:primas}
  }
\end{figure}
Following this observation, the structure of 
a series of three-particle systems was studied in Ref.~\cite{MaHuMuRe11a} 
without imposing any \emph{a priori} 
separation or assumption about the motion of the different particles.
At the same time, concerning our original observation, 
the lack of any definition of molecular structure beyond
the BO approximation, we may wonder how to study the structure of a molecule 
in a theoretical framework which does not invoke the BO approximation? 

Observables in quantum mechanics are computed as the expectation value of 
the appropriate operator with the wave function of the system. 
However, if we calculated the carbon nucleus-proton distance in an organic molecule, 
we would obtain a single $\langle \Psi | r_{\text{CH}} \Psi\rangle$ value \cite{CaAd04,SuWo05,SuWoCPL05}, 
irrespective of which proton and which carbon we pick, 
due to the quantum mechanical indistinguishability of identical particles. 
To give another example, the single expectation value of the HHH angle in H$_3^+$ 
is not useful to distinguish between the linear and triangular arrangements of the three protons, 
since the expectation, \emph{i.e.,} average, 
$\langle\alpha_\mr{HHH}\rangle=\langle \Psi | \alpha_{\text{HHH}} \Psi\rangle$ value is 
the same for either a linear
$\langle\alpha_\mr{HHH}\rangle=(0^\mr{o}+180^\mr{o}+0^\mr{o})/3=60^\mr{o}$ or
a triangular arrangement, $\langle\alpha_\mr{HHH}\rangle=(60^\mr{o}+60^\mr{o}+60^\mr{o})/3=60^\mr{o}$.
The general problem of the reconciliation of the classical molecular structure theory with 
a full many-particle quantum description has been recognized decades ago and was 
also referred to as the \emph{molecular structure conundrum} \cite{We84} 
(further relevant references include \cite{Wo76,ClDi80,To81,We84}).

\subsection{Probabilistic interpretation of the wave function}
Claverie and Diner suggested in 1980 
that appropriate marginal probability density functions calculated from 
the full wave function could be used to identify molecular structural features 
in the full electron-nuclear wave function \cite{ClDi80}. 
In other words, structural parameters do not have sharp, 
dispersionless values, but they are characterized by some probability density function.
This idea has been explored for 
the analytically solvable Hooke--Calogero model of molecules \cite{UMH06,UMH08,LuEcLoUg12,BePoLu}.
The atoms-in-molecule analysis has been extended to the realm of electron-nuclear quantum theory 
\cite{GoSh11,GoSh12}. Most recently, 
it was demonstrated that the proton density in methanol obtained from 
an electron-proton orbital computation (with fixed carbon and oxygen nuclei) 
can be matched with the spatial configuration obtained 
from a BO electron-structure calculation \cite{KaYa12}. Furthermore, in addition
to the electronic and nuclear densities, flux densities have also been considered
in Refs.~\cite{PeTo13,PeTo15,PoTr16}.

For the sake of the present discussion, we shall stay with the analysis of (molecular) structure
in terms of probability density functions calculated from the full wave function. Further 
general discussion of obtaining the classical molecular structure from 
quantum mechanics is provided in Section~\ref{ch:classquant}.
In what follows, one- and two-particle probability density 
functions \cite{MaHuMuRe11a,MaHuMuRe11b}
are introduced which will be used for the structural analysis later in this section.
The probability density of selected particles measured 
from a ``center point'' $\cp$ fixed to the body is 
  \begin{align}
    &D^{(n)}_{\cp,a_1 a_2 \ldots a_n}(\mx{R}_1,\mx{R}_2,\ldots,\mx{R}_{n})
      \nonumber \\
    &=
    \langle %
      \Psi |
      \delta(\mx{r}_{a_1}-\mx{r}_{\cp} - \mx{R}_1)
      \delta(\mx{r}_{a_2}-\mx{r}_{\cp} - \mx{R}_2)
      \ldots
      \delta(\mx{r}_{a_n}-\mx{r}_{\cp} - \mx{R}_n)
      | \Psi
    \rangle
  \end{align}
  with $\mx{R}_i\in\mathbb{R}^3$ and the three-dimensional Dirac delta distribution, $\delta(\mx{r})$.
  The center point $\cp$ can be the center of mass (denoted by ``0'') or another particle.
  For a single particle, this density function is
  \begin{align}
    D_{\cp,a}^{(1)}(\mx{R}_1) =
    \left\langle %
      \Psi |\delta(\mx{r}_a-\mx{r}_{\cp}-\mx{R}_1)| \Psi
    \right\rangle.
    \label{eq:d0a}
  \end{align}
  For $\cp=0$, $D_{0,a}^{(1)}$ is the spatial density
  of particle $a$ around the center of mass (``0''),
  while for $\cp=b$, $D_{b,a}^{(1)}$
  measures the probability density of the displacement vector connecting 
  $a$ and $b$.
  
  Due to the overall space rotation-inversion symmetry, 
  $D_{\cp,a}^{(1)}(\mx{R}_1)$ is ``round'' for $N=0$, $p=+1$ and
  the corresponding radial function is:
  \begin{align}
    \rho_{P,a}(R)=D_{\cp,a}^{(1)}(\mx{R}_1)
    \label{eq:rho}
  \end{align}
  with $\mx{R}_1=(0,0,R)$ and $R\in\mathbb{R}^+_0$. 
  We normalize the density functions to one 
  (so, they measure the fraction of particles which can be found in 
  an infinitesimally small interval $\text{d}R$ around $R$):
  \begin{align}
    4\pi \int_0^\infty \dd R\ R^2\ \rho_{\cp,a}(R) = 1.
  \end{align}
  The probability density function for the included angle $a$--$P$--$b$
  is obtained by integrating out the radii in 
  the two-particle density measured from a center point $\cp$
  \begin{align}
    \iGamma_{\cp,ab}(\alpha) 
    = 
    \int_0^\infty \dd R_1 R_1^2\ 
    \int_0^\infty \dd R_2 R_2^2\ 
      D_{\cp,ab}^{(2)}(\mx{R}_1,\mx{R}_2) ,
    \label{eq:defGam}
  \end{align}
  with
  \begin{align}
    D^{(2)}_{\cp,ab}(\mx{R}_1,\mx{R}_2) =
    \left\langle %
      \Psi|
       \delta(\mx{r}_a-\mx{r}_{\cp}-\mx{R}_1)
       \delta(\mx{r}_b-\mx{r}_{\cp}-\mx{R}_2)
      |\Psi
    \right\rangle .
    \label{eq:d2cacb}
  \end{align}
The center point, $\cp$, can be the center of mass ($\cp=0$) or another particle ($\cp=c$).
Similarly to $D_{\cp,a}^{(1)}(\mx{R}_1)$,
$D_{\cp,ab}^{(2)}(\mx{R}_1,\mx{R}_2)$ is also spherically symmetric for 
wave functions with $N=0$, $p=+1$, and its numerical value depends only on 
the lengths $R_1=|\mx{R}_1|$, $R_2=|\mx{R}_2|$, and the $\alpha$ included angle of 
the vectors $\mx{R}_1$ and $\mx{R}_2$ (for non-zero lengths). 
We normalize the angle density according to 
  \begin{align}
    8\pi^2 \int_0^\pi \dd\alpha\sin\alpha\ \iGamma_{\cp,ab}(\alpha) = 1.
  \end{align}

\subsubsection{Numerical demonstration of the H$^-\longrightarrow$\ H$_2^+$ transition}
Following Hans Primas' observation (Figure~\ref{fig:primas})
Ref.~\cite{MaHuMuRe11a} studied 
the family of $\lbrace a^\pm,a^\pm,b^\mp \rbrace$-type
three-particle Coulomb interacting systems with 
two identical particles and a third one. 
This family of systems is described by the Hamiltonian
\begin{align}
  \hat{H}(\mapart,\mbpart,\mx{r}) 
   =  
   - \frac{1}{2\mapart}\Delta_{\mx{r}_1}
   - \frac{1}{2\mapart}\Delta_{\mx{r}_2}
   - \frac{1}{2\mbpart}\Delta_{\mx{r}_3} 
   + \frac{1}{|\mx{r}_1-\mx{r}_2|}
   - \frac{1}{|\mx{r}_1-\mx{r}_3|}
   - \frac{1}{|\mx{r}_2-\mx{r}_3|},
   \label{eq:aabham}
\end{align}
for various masses and unit charges. Note that the Hamiltonian 
is invariant to the inversion of the electric charges. 
Furthermore, rescaling the masses by a factor 
$\eta$ is equivalent to scaling the energy and shrinking the length by the factor $\eta$   
\begin{align}
  \hat{H}(\eta m_a,\eta m_b,\mx{r})= \eta \hat{H}(m_a,m_b,\eta\mx{r}),
  \quad\quad \forall\eta\in\mathbb{R}\setminus\{0\} .
\end{align}  
Thereby, it is sufficient to consider only the $\massratio$ mass ratio to obtain 
qualitatively different solutions.
It is also known that the Hamiltonian in Eq.~(\ref{eq:aabham}) has at least 
one bound state for all mass ratios \cite{MaRiWu92,FaEsKa99}. 

\begin{figure}
  \begin{center}
    \includegraphics[scale=1.]{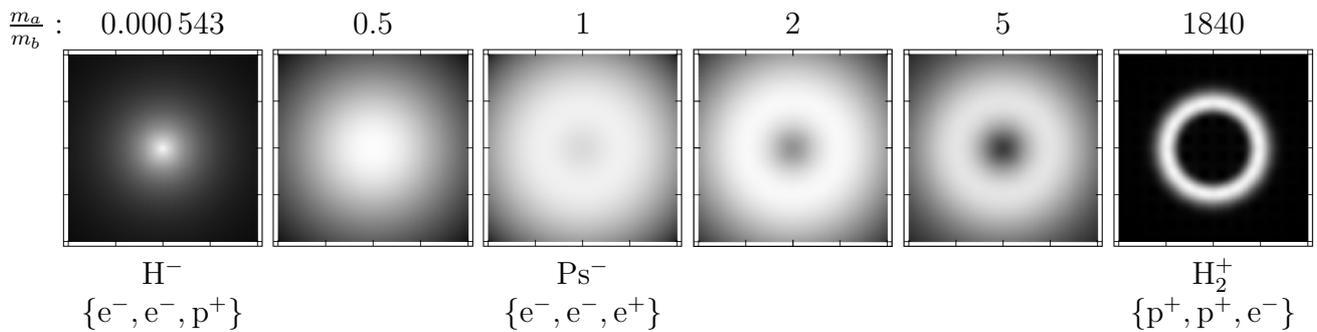}
  \end{center}
  \caption{%
    Transition of the ground-state particle density, $D^{(1)}_{0a}$, 
    by increasing the $\massratio$ mass ratio in $\lbrace a^\pm,a^\pm,b^\mp\rbrace$-type
    systems \cite{MaHuMuRe11a}. The center (0) of each plot is the center of mass.
    \label{fig:emer}
  }
\end{figure}

To numerically study the H$^-\rightarrow\ \ $H$_2^+$ transition, 
the ground-state wave functions were computed \cite{MaHuMuRe11a} 
for several mass ratio values using the variational procedure 
described in Section~\ref{ch:varsol}. 
Figure~\ref{fig:emer} shows the transition of the particle density, $D^{(1)}_{0a}$. 
It is interesting to note that 
the emergence of the particle shell is solely induced by the increase of 
$\massratio$ \cite{MaHuMuRe11a,MaHuMuRe11b}, while the symmetry-properties of 
the systems remain unchanged. 
All systems are ``round'' in their ground state with $N=0$ and $p=+1$. 
In Ref.~\cite{MaHuMuRe11a} the transition point was numerically estimated 
to be between 0.4 and 0.8,
which also suggests that the positronium anion, 
Ps$^-=\lbrace\text{e}^-,\text{e}^-,\text{e}^+\rbrace$, has some molecular character. 
The figure represents the H$_2^+$ molecular ion as a shell. 
We may wonder whether it is possible to identify the relative position of 
the protons within the shell. For this purpose
the angular density function, $\Gamma_{0,pp'}$, was
calculated in Ref.~\cite{MaHuMuRe11b}, 
which demonstrated that the protons are indeed found at around antipodal points 
of the shell (remember that the center of each plot is the center of mass). 
As to earlier theoretical work, 
Kinsey and Fröman \cite{FrKi61} and later Woolley \cite{Wo76} have anticipated 
similar results by considering the ``mass polarization'' 
in the translationally invariant Hamiltonian 
arising due to the separation of the center of mass. 
Furthermore, the proton shell has some finite width, 
which can be interpreted as the zero-point vibration in the BO picture. 
Recent work \cite{KiLoCo13,KiRhCo16,BaKiCo16} has elaborated more 
on the transition properties and vibrational dynamics of this family of 
three-particle systems and determined the mass ratio where 
the transition takes place more accurately.

\subsubsection{Numerical example for a triangular molecule}
\begin{figure}
  \begin{center}
    \includegraphics[scale=1.]{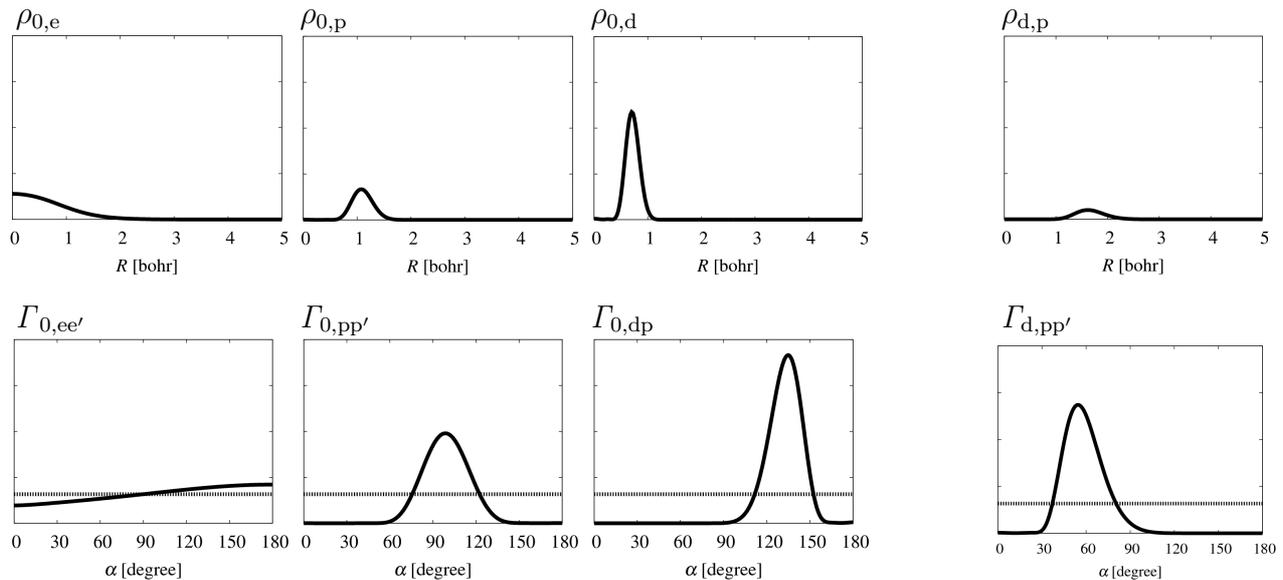}
  \end{center}
  \caption{%
    Radial, $\rho_{ab}$, and angular, $\iGamma_{a,bc}$,
    probability density functions computed for
    H$_2$D$^+=\{\mr{e}^-,\mr{e}^-,\mr{p}^+,\mr{p}^+,\mr{d}^+\}$.
    \label{fig:h2dp}
  }
\end{figure}

In the particle density plots, 
larger molecules would also be seen as ``round'' objects 
in their eigenstates with zero total angular momentum and 
positive parity ($N=0,p=+1$), and localized particles 
form shells around the molecular center of mass. 
In order to demonstrate a non-trivial 
arrangement of the atomic nuclei within a molecule, the H$_2$D$^+=\lbrace
\text{p}^+,\text{p}^+,\text{d}^+,\text{e}^-,\text{e}^- \rbrace$ molecular ion
was studied in Ref.~\cite{MaHuMuRe11b}.
Interestingly, the qualitative features of the computed density functions 
(see Figure~\ref{fig:h2dp}) converged very fast, small basis sets and 
a loose parameterization was sufficient to observe converged structural features, 
whereas the energies were far from spectroscopic accuracy. 

Figure~\ref{fig:h2dp} summarizes the particle-density functions 
which highlight characteristic structural features of the system. 
First, we can observe the delocalized electron cloud ($\rho_{0,\text{e}}$), 
the proton shell ($\rho_{0,\text{p}}$), and the deuteron shell ($\rho_{0,\text{d}}$) 
around the center of mass. The deuteron shell is more peaked and more localized 
in comparison with the proton shell. 
(Remember that these plots show the spherically symmetric density 
along a ray, $\rho_{ab}$, and all density functions are normalized to one.) 

Next, let's look at the probability density functions for the included angle, $\iGamma_{0,ab}$, 
of two particles measured from the molecular center of mass (``0''). 
The dashed line in the plots shows the angular density corresponding to a hypothetical system 
in which the two particles ($a$ and $b$) are independent. 
It is interesting to note that for the two electrons $\iGamma_{0,\text{ee'}}$ shows 
very small deviation from the (uncorrelated system's) dashed line. 
At the same time,
we see a pronounced deviation from the dashed line 
for the nuclei, $\iGamma_{0,\text{pp'}}$ and $\iGamma_{0,\text{pd}}$. %
These numerical observations are in line with 
Claverie and Diner's suggestion based on theoretical considerations \cite{ClDi80} that 
molecular structure could be seen in an fully quantum-mechanical description 
as correlation effects for the nuclei. 

As to the included angle of the two protons and the deuteron,
the $\iGamma_{\text{d},\text{pp'}}$ probability density function 
has a maximum at around 60 degrees, which indicates the triangular arrangement of the nuclei.
Due to the almost negligible amplitude of $\iGamma_{\text{d},\text{pp'}}$ at around 180 degrees
the linear arrangement of the three nuclei (in the ground state) can be excluded.
Thus, the structure of H$_2$D$^+$ derived from our pre-BO numerical study is in agreement
with the equilibrium structure known 
from BO electronic-structure computations.

\subsection{Classical structure from quantum mechanics \label{ch:classquant}}
\begin{figure}
  \includegraphics[scale=1.]{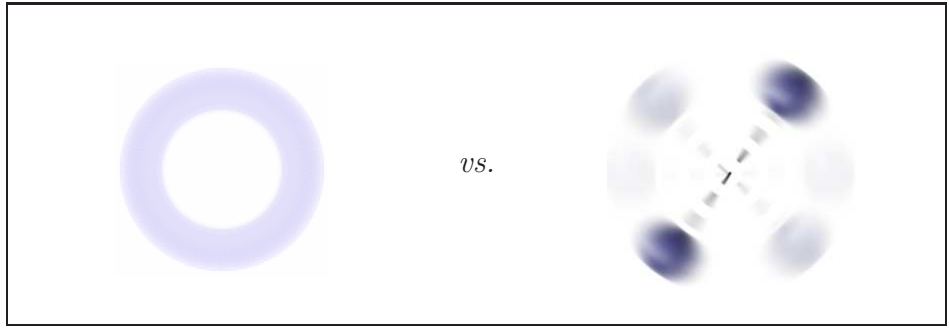}  
  \caption{%
    Quantum vs. classical structure of molecules: 
    superposition or rotating dumbbell.
    \label{fig:palcika}
  }
\end{figure}

Relying on the probabilistic interpretation of quantum mechanics 
the structure of 
H$_2^+$ was visualized as a proton shell (Figure~\ref{fig:emer}) 
with the protons found at around the antipodal points,
and 
H$_2$D$^+$ as a proton shell and 
a deuteron shell within which the relative arrangement of the three nuclei is dominated 
by a triangle (Figure~\ref{fig:h2dp}). This analysis has demonstrated
that \emph{elements} of molecular structure can be \emph{recognized} in
the appropriate marginal probability densities calculated 
from the full electron-nuclear wave function. 
At the same time, a chemist would rather think about H$_2^+$
as a (classical) rotating dumbbell (Figure~\ref{fig:palcika})
and H$_2$D$^+$ as a (nearly) equilateral triangle. 
Although \emph{elements} can be recognized in the probability density functions, 
the link to the classical structure which 
chemists have used for more than a century to understand and design 
new reaction pathways for new materials, 
is not obvious \cite{Hu27,Primas81,Wo76,CaCh98,SuWo05,SuWoCPL05}.
In order to recover the classical molecular structure from a 
fully quantum mechanical treatment, it is necessary to obtain for a molecule
\begin{itemize}
  \item[(a)]
    the shape;
  \item[(b)]  
    the handedness: chiral molecules are found exclusively in their left- or right-handed 
    version or a classical mixture (called racemic mixture) of these mirror images 
    but ``never'' in their superposition;
  \item[(c)]
    the individual labelling of the atomic nuclei (distinguishability).
\end{itemize}
Although it is possible to write down appropriate linear combinations (wave packets) of 
eigenstates of the full Hamiltonian, which satisfy these requirements at certain moments, 
we would like to recover these properties as permanent \emph{molecular observables}.

A possible resolution of this puzzle would be the description of 
the molecule as an open quantum system being in interaction with 
an environment \cite{JoZe03,Schloss07}. According to decoherence theory pointer states 
are selected by the continuous monitoring of the environment. 
As a result, the system's reduced density matrix 
(after tracing out the environmental degrees of freedom from the world's density matrix) 
written in this pointer basis evolves 
in time so that its off-diagonal elements decay exponentially with some decoherence 
time characteristic to the underlying microscopic interaction process 
with the environment (radiation or matter). This decay of the off-diagonal elements leads 
to the suppression of the interference terms between different pointer states, 
and results in a (reduced) density matrix the form of which corresponds to that of mixed states. 
Hence, this result can be \emph{interpreted} as the emergence of the classical features 
in a quantum mechanical treatment 
\footnote{There is an unsettled discussion concerning the mixed states of open quantum systems 
in terms of proper vs. improper mixtures, which is related to 
the quantum measurement problem \cite{FoLoGo16,Ba16}.}. 
So, decoherence theory allows 
us to identify pointer states, which are selected and remain stable as a result of 
the molecule's interaction with its environment. 

It is interesting to note that important molecular properties 
(shape, handedness, atomic labels) break the fundamental symmetries of 
an isolated quantum system:
the rotational and inversion symmetry, 
as well as the indistinguishability of identical particles. 
It remains a task to explore on a detailed microscopic level 
how and why these broken-symmetry states become pointer states of a molecular system.

\paragraph{Shape}
Following the pioneering studies which have identified pointer states and confirmed 
their stability upon translational localization \cite{HoSi03,Ad06,BuHo09}
Ref.~\cite{ZhRo16} provides a detailed account of the rotational decoherence of 
mesoscopic objects induced by a photon-gas environment or massive particles 
in thermal equilibrium. The qualitative conclusions are similar for 
the two different environments, but there are differences in the estimated 
decoherence time and its temperature dependence differ for the two environments. 
Orientational localization of the mesoscopic ellipsoid takes place only 
if there are at least two directions for which the electric polarizabilities are different, 
and coherence is suppressed exponentially with the angular distance between two orientations.

\paragraph{Handedness}
As to the chirality of molecules, the superselection phenomenon has been demonstrated in 
Ref.~\cite{TrHo09} by using a master equation \cite{Ho07} which describes 
the incoherent dynamics of the molecular state in the presence of the scattering of 
a lighter, thermalized background gas. Experimental conditions are predicted under 
which the tunneling dynamics is suppressed between the left and right-handed 
configurations of D$_2$S$_2$.  
%
%

\paragraph{Individual labelling of the atomic nuclei}
Concerning the distinguishability of atomic nuclei, it remains a challenge to work out 
the detailed theoretical equations and to estimate the experimental conditions under which 
the individual labelling of quantum mechanically identical atomic nuclei 
(\emph{e.g.,} protons) emerges.

%
%
\section{Summary and future challenges\label{ch:future}} 
The direct solution of the full electron-nuclear Schr\"odinger equation, 
without the introduction of any kind of separation of the electronic and the nuclear motion, 
makes it possible to approach the non-relativistic limit arbitrarily close. 
It is interesting to note that rovibrational states bound by 
an excited electronic state within the Born--Oppenheimer approximation 
are obtained as resonances within a pre-Born--Oppenheimer treatment giving access 
to not only the energy but also the finite predissociation lifetime of the state.
Although the variational principle offers a simple strategy to assess and systematically 
improve the accuracy of eigenvalues and eigenfunctions of bound states, 
a systematic variational computation of unbound states is more challenging.
In short, it is an outstanding challenge to 
develop efficient computational algorithms 
which give access with sub-spectroscopic accuracy to
the ground and excited, bound and unbound rovibronic states of polyatomic molecules.
%
These developments will serve as solid basis for further 
work outlined in the following directions.

\paragraph{Theoretical developments for precision spectroscopy of polyatomic molecules}
The comparison of highly accurate molecular computations and 
precision measurements 
contributes to the testing of fundamental physical constants, \emph{e.g.,}
the proton-to-electron mass ratio, the fine structure constant, 
or to pinpoint fundamental physical quantities, such as the proton radius, 
and to test fundamental physical theories \cite{metro16,UbKoEiSa16}. 
For a meaningful comparison it is mandatory to be able to solve the non-relativistic
Schrödinger equation very accurately and to account for the relativistic as well as 
quantum electrodynamics (QED) effects.

\paragraph{Hierarchy of approximate pre-BO methods}
The idea of including the atomic nuclei in the quantum mechanical treatment of the electrons
has been pursued in order to develop a systematically improvable, 
hierarchy of approximate pre-BO methods 
\cite{Nakai02,BoVaSh04,ChPaHS08,IsTaNa09,SiPaSwHS13,Lara13,lowdin13}.
An appealing feature of a pre-BO treatment is that it obviates the need for the computation and 
fitting of the potential energy surface(s) and 
non-adiabatic coupling vectors (for multiple electronic states).
At the moment, it appears to be technically and computationally extremely challenging to devise 
a practical, accurate, and systematically improvable hierarchy of approximate electron-nuclear 
orbital methods.
Recently,
a combination of electronic structure and quantum nuclear motion theory
has been suggested \cite{CaChSuLi15,CaChSuLi17}, which aims to combine the
best of the two worlds in a practical approach.

\paragraph{Towards a molecular decoherence theory}
The definition of molecular structure within a fully quantum mechanical description of 
molecules remains to be an unsettled problem \cite{Hu27,Primas81,Wo76,CaCh98,SuWo05,SuWoCPL05} 
either for a numerically ``exact'' or an approximate treatment. 
Certainly, the probabilistic interpretation of the molecular wave function and 
the study of appropriate marginal probability densities provide 
useful pieces of information about the structure of a molecule. 
In order to arrive at a quantum molecular theory, 
in which the molecule is treated as a whole quantum mechanically,
and at the same time
the known chemical concepts are restored from the theoretical treatment, 
it is necessary to re-establish the shape, the handedness, and 
the individual labeling of identical atomic nuclei from the quantum mechanical description.
Interestingly, these important chemical properties break 
the fundamental symmetries of an isolated quantum system.
The application of decoherence theory with realistic microscopic models for molecules offers 
a reasonable starting point for the reconstruction of these known classical chemical properties. 
The estimation of decoherence time for various environments and interactions has relevance
for the practical realization of quantum control and quantum computing experiments with molecules.

\vspace{1cm}
\noindent %
\textbf{Acknowledgment} \\
Financial support from a PROMYS Grant (no. IZ11Z0\_166525)  
of the Swiss National Science Foundation is gratefully acknowledged.

\clearpage

\begin{thebibliography}{109}
\bibitem[{spe()}]{spectracc}
\bibinfo{note}{The term ``spectroscopic accuracy'' is not uniquely defined but
  it is usually used to refer to computations providing vibrational transition
  wave numbers with a certainty better than 1~\cm\ ($\approx 4.6\
  \mu$E$_\mr{h}$) and even tighter accuracy for rotational transitions.}
\end{thebibliography}

\end{document}